\documentclass[]{natureprintstyle}
\usepackage{multirow}
\usepackage{savesym}
\usepackage{amsmath}
\savesymbol{iint}
\usepackage{txfonts}
\restoresymbol{TXF}{iint}
\usepackage{graphicx}
\usepackage{ulem} 
\usepackage{url}
\usepackage{threeparttable}
\usepackage{blindtext}
\usepackage{xcolor}
\usepackage{soul}
\usepackage{booktabs}
\usepackage[super,sort,comma]{natbib}
\usepackage{bibunits}

\usepackage{multirow}
\usepackage{savesym}
\usepackage{amsmath}
\savesymbol{iint}
\usepackage{txfonts}
\restoresymbol{TXF}{iint}
\usepackage{graphicx}
\usepackage{ulem}
\usepackage{url}
\usepackage{threeparttable}
\usepackage{blindtext}
\usepackage{xcolor}
\usepackage{soul}
\usepackage{booktabs}
\usepackage[super,sort,comma]{natbib}
\usepackage{bibunits}

\usepackage{etoolbox}
\makeatletter
\newcommand*{\newbibstartnumber}[1]{%
  \apptocmd{\thebibliography}{%
    \global\c@NAT@ctr #1\relax
    \addtocounter{NAT@ctr}{-1}%
  }{}{}%
}
\makeatother

\usepackage[colorlinks, linkcolor=violet, anchorcolor=green, citecolor=blue]{}

\def\urlprefix   {{\sc url: }}

\def\twco {$^{12}$CO}%
\def\thC {$^{13}$C}%
\def\etO {$^{18}$O}%
\def\new#1 {{\bf #1 }}

\def\purple#1 {{\textcolor{purple}{#1}}\ }
\def\red#1 {\textcolor{red}{#1}}
\def\new#1 {{\bf #1 }}
\def\emph#1  {\textit{ #1 } }%

\newcommand{\apj}{Astrophys. J.}

\newcommand{\pasp}{Publ. Astron. Soc. Pac.}
\newcommand{\apjs}{Astrophys. J. Supp.}
\newcommand{\araa}{Annu. Rev. Astron. Astrophys.}
\newcommand{\mnras}{Mon. Not. R. Astron. Soc.}
\newcommand{\apjl}{Astrophys. J. Let.}
\newcommand{\aap}{Astron. Astrophys.}

\newcommand{\nat}{Nature}

\newcommand{\aaps}{A\&AS}
\newcommand{\aapr}{A\&ARv}
\newcommand{\fcp}{Fundamentals of Cosmic Physics}

\title{Stellar populations dominated by massive stars in dusty starburst galaxies across cosmic time}

\author{Zhi-Yu~Zhang$^{1,2}$, D.~Romano$^3$, R.\,J.~Ivison$^{2,1}$, P.\,P.~Papadopoulos$^{4,5}$ \& F.~Matteucci$^{6,7,8}$}

\begin{document}

\begin{bibunit}[plainnat]

\maketitle

\begin{affiliations}
\item Institute for Astronomy, University of Edinburgh, Royal Observatory, Blackford Hill, Edinburgh EH9 3HJ, UK
\item ESO, Karl-Schwarzschild-Str.~2, D-85748 Garching, Germany
\item INAF, Astrophysics and Space Science Observatory, Bologna via Piero Gobetti 93/3, 40129, Bologna, Italy 
\item Department of Physics, Section of Astrophysics, Astronomy and Mechanics, Aristotle University of Thessaloniki, Thessaloniki 54124, Greece
\item Research Center for Astronomy, Academy of Athens, Soranou Efesiou 4, GR-115 27 Athens, Greece
\item Department of Physics, Section of Astronomy, University of Trieste, Trieste, Italy
\item INAF, Osservatorio Astronomico di Trieste, Via Tiepolo 11, I-34131 Trieste, Italy
\item INFN, Sezione di Trieste, Via Valerio 2, I-34127 Trieste, Italy
\end{affiliations}


\begin{abstract}
All measurements of cosmic star formation must assume an initial distribution
of stellar masses -- the stellar initial mass function -- in order to
extrapolate from the star-formation rate measured for typically rare, massive
stars ($M_\star \ge\ 8$ M$_{\odot}$) to the total star-formation rate across
the full stellar mass spectrum\cite{Kennicutt1998}.  The shape of the stellar
initial mass function in various galaxy populations underpins our understanding
of the formation and evolution of galaxies across cosmic
time\cite{Bastian2010}. Classical determinations of the stellar initial mass
function in local galaxies are traditionally made at ultraviolet, optical and
near-infrared wavelengths, which cannot be probed in dust-obscured
galaxies\cite{Kroupa2013,Bastian2010}, especially in distant starbursts, whose
apparent star-formation rates are hundreds to thousands of times higher than in
our Milky Way, selected at submillimetre (rest-frame far-infrared)
wavelengths\cite{Smail1997,Hughes1998}. The $^{13}$C/$^{18}$O\ abundance ratio
in the cold molecular gas -- which can be probed via the rotational transitions
of the $^{13}$CO and C$^{18}$O isotopologues -- is a very sensitive index of
the stellar initial mass function, with its determination immune to the
pernicious effects of dust. Here we report observations of $^{13}$CO and
C$^{18}$O emission for a sample of four dust-enshrouded starbursts at redshifts
of approximately two to three, and find unambiguous evidence for a top-heavy
stellar initial mass function in all of them. A low
$^{13}$CO/C$^{18}$O\ ratio for all our targets -- alongside a well-tested,
detailed chemical evolution model benchmarked on the Milky Way\cite{Romano2017}
-- implies that there are considerably more massive stars in starburst events
than in ordinary star-forming spiral galaxies. This can bring these
extraordinary starbursts closer to the `main sequence' of star-forming
galaxies\cite{Noeske2007}, though such main-sequence galaxies may not be immune
to changes in initial stellar mass function, depending upon their
star-formation densities. 

\end{abstract}

Oxygen, carbon and their stable isotopes are produced solely by nucleosynthesis in
stars\cite{Wilson1992}. The minor isotopes, $^{13}$C\ and $^{18}$O, are
released mainly by low- and intermediate-mass stars (LIMSs; $M_{\star}$\
$<$ 8 M$_{\odot}$) and massive stars ($M_{\star}$\ $>$ 8 M$_{\odot}$),
respectively\cite{Romano2010}, owing to their differing energy barriers in
nuclear reactions and evolution of stars\cite{Pagel2009}. These isotopes then
mix with the interstellar medium (ISM) such that the $^{13}$C/$^{18}$O\
abundance ratio measured in ISM becomes a `fossil', imprinted by evolutionary
history and the stellar initial mass function (IMF)\cite{Romano2017}. The
abundances of the $^{13}$CO and C$^{18}$O isotopologues in the molecular ISM,
whose measurements are immune to the pernicious effects of dust, are therefore
a very sensitive index of the IMF in galaxies.

Galaxies in the early Universe, having had much less cosmological time
available for prior episodes of evolution, are expected to have simpler
star-formation histories than local galaxies. Our sample comprises the
strongest carbon monoxide (CO) emitters in the early Universe, selected from
the literature (see Methods): four gravitational lensed submillimetre galaxies
(SMGs) at $z\sim2$--3, with look-back times, $\gtrsim 10$\,Gyr.

Using the Atacama Large Millimeter Array (ALMA), we have robustly ($> 5 \sigma$,
where $\sigma$ is the standard deviation) detected multiple transitions of
$^{13}$CO\ and C$^{18}$O\ in most of our target galaxies. The $J=3\rightarrow2$
line from SDP.17b and the $J=5\rightarrow4$ line from SPT\, 0103$-$45 are
marginally detected at $\sim 4 \sigma$ levels. But the $J=4\rightarrow3$
transitions of SDP.17b are detected at high signal-to-noise so we can be
confident that emission features seen at the expected velocities of the weaker
transitions are also real. We also detected \twco\ $J= 4 \rightarrow3$
and $J= 5 \rightarrow4$ for SPT 0125$-$47 and SPT 0103$-$45, respectively.

\begin{figure}
\centering
\includegraphics[scale=1.05]{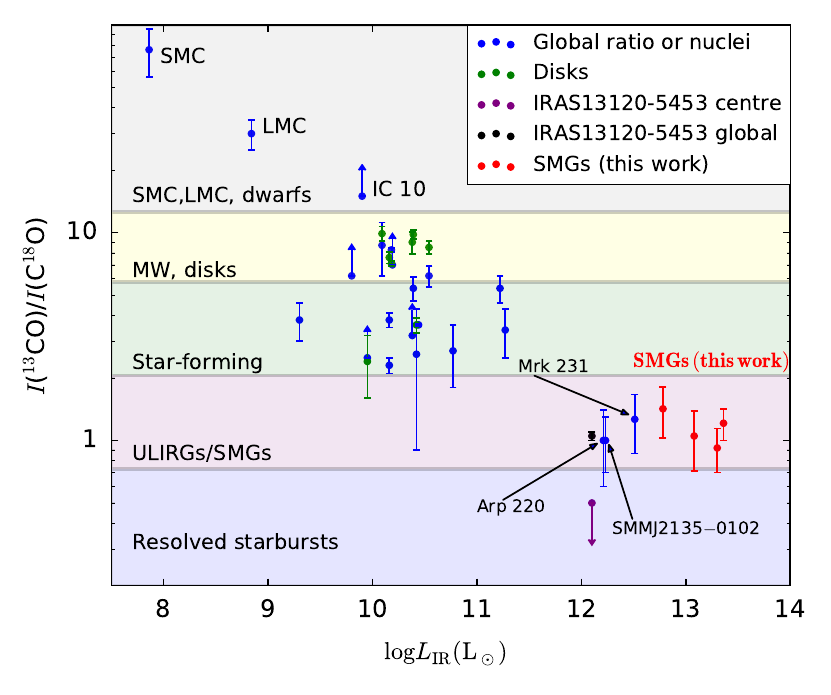}
\caption{$I$($^{13}$CO)/$I$(C$^{18}$O) as a function of IR luminosity ( $L_{\rm
IR}$, rest-frame 8--1000\,$\mu$m, corrected for gravitational amplification
when appropriate). Red symbols refer to submillimetre galaxies (SMGs) in our
sample. We include $I$($^{13}$CO)/$I$(C$^{18}$O) measurements of nearby
star-forming spiral galaxies, disks, a few resolved nuclei\cite{JD2017}, three
local ULIRGs\cite{Henkel2014}, Arp\,220,  Mrk\,231, and IRAS\,13120$-$5453, and
a SMG, SMM\,J2135$-$0102 \cite{Danielson2013} at $z\approx 2.3$. The ratios of
the Small and Large Magellanic Clouds (SMC and LMC) are averaged from multiple
positions \cite{Heikkila1998,Muraoka2017}.  $I$($^{13}$CO)/$I$(C$^{18}$O) in a
dwarf galaxy IC\,10 is reported as a lower limit\cite{Nishimura2016}. A
decreasing trend of $I$($^{13}$CO)/$I$(C$^{18}$O) as a function of $L_{\rm
IR}$\ is clearly evident, indicating that $^{13}$CO/C$^{18}$O\ abundance ratios
are varying systematically in galaxies with different rates of apparent star
formation. Error bars stand for an one $\sigma$ uncertainty.}
\label{fig:ratios}
\end{figure}

As shown in Fig.~\ref{fig:ratios}, there is a decreasing trend in the ratio of
velocity-integrated line intensities, $I$($^{13}$CO)/$I$(C$^{18}$O), with
increasing infrared luminosity, $L_{\rm IR}$ (or, the apparent star-formation
rate, SFR, traced by massive stars). For all the galaxies in our observed
sample (see Methods), the line ratios of $I$($^{13}$CO)/$I$(C$^{18}$O) are
close to unity, similar to those found\cite{Henkel2014,Sliwa2017} in three
nearby ultraluminous infrared galaxies (ULIRGs; $L_{\rm IR}> 10^{12}$
L$_{\odot}$) -- Arp\,220, Mrk\,231 and IRAS\,13120$-$5453, as well as in the
strongly lensed SMG, SMM\,J2135$-$0102 at $z\approx 2.3 $\cite{Danielson2013}.
Galactic disks of nearby spiral galaxies have $I$($^{13}$CO)/$I$(C$^{18}$O)
ratios similar to the representative ratio\cite{Wilson1992,Barnes2015} of our
Milky Way's disk, $\sim 7$--10. In the central nuclear regions of these spiral
galaxies, where the star-formation activity is more intense than in the disks,
$I$($^{13}$CO)/$I$(C$^{18}$O) ratios are lower\cite{JD2017}, though remain
restricted to $I$($^{13}$CO)/$I$(C$^{18}$O) $\ge 4$.  The Magellanic clouds --
our nearest dwarf galaxies -- show the highest $I$($^{13}$CO)/$I$(C$^{18}$O)
ratios, $\gtrsim 30$--60.

For representative Galactic abundance ratios of $^{13}$CO/C$^{18}$O\ $\sim
7$--10, the $I$($^{13}$CO)/$I$(C$^{18}$O) line intensity ratio can approach
values near unity -- which is what we measure for all the galaxies in our
sample -- only if even the rarest of our three isotopologue lines, C$^{18}$O,
were to acquire substantial optical depths on galactic scales (see Methods). On
the other hand, to reach line ratios, $I$($^{12}$CO)/$I$($^{13}$CO) and
$I$($^{12}$CO)/$I$(C$^{18}$O) in excess of 30 -- as found in our sample -- the
optical depth of $^{13}$CO\ and C$^{18}$O have to be $\ll 1$ for either type of
conditions, namely local thermodynamic equilibrium (LTE) or non-LTE excitation
(see Methods), assuming the typical abundance ratios of $^{12}$CO/$^{13}$CO\
$\sim 40$--100 found in the Milky Way\cite{Wilson1992,Romano2017}.

The magnification factors of gravitational lensing in our objects are modest
($\mu \sim 5$, see Extended Data Table~\ref{tab:targets}), with the notable
exception of the Cloverleaf ($\mu \sim 10$). It is unlikely that differential
lensing could skew the measured $I$($^{13}$CO)/$I$(C$^{18}$O) line ratio away
from the value intrinsic to the galaxy, even in cases of much stronger
lensing\cite{Danielson2013}. For differential lensing to operate in this way,
the global $I$($^{13}$CO) and $I$(C$^{18}$O) distributions over the galaxies
must be very different -- which is improbable given that these two isotopologue
lines have almost identical excitation requirements and any differences in
their distribution are expected to be confined within individual molecular
clouds (see Methods). Finally, the isotopologue lines have been observed
simultaneously, making the uncertainties from pointing and calibration
negligible. Furthermore, it was recently shown that known photo-chemical
effects, such as selective photodissociation and fractionation, cannot induce
global isotopologue abundances to differ from the intrinsic, IMF-determined,
isotopic abundances in star-forming galaxies\cite{Romano2017,Sliwa2017}.

\begin{figure}
\centering
\includegraphics[scale=1.05]{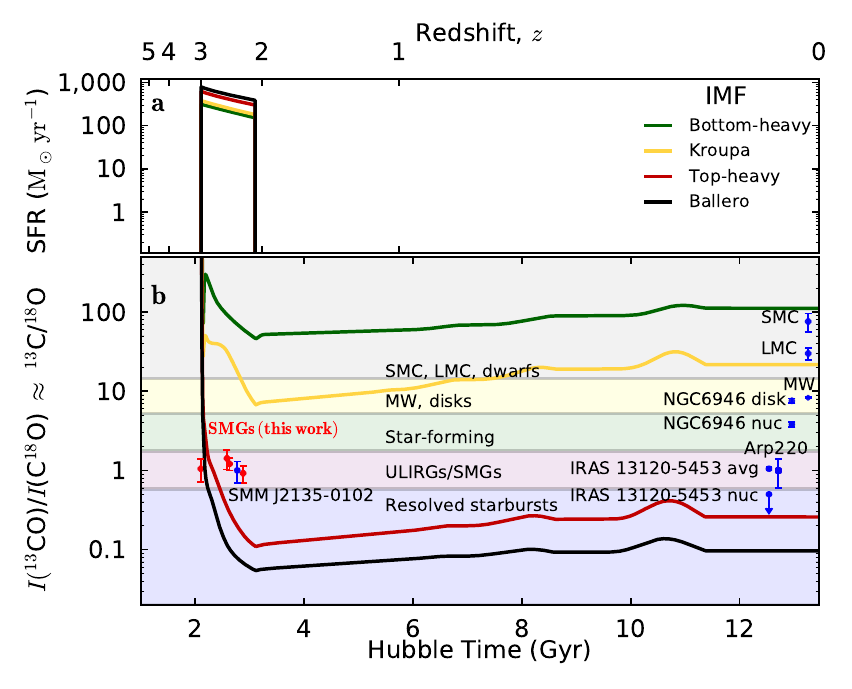}
\caption{Theoretical \thC\ and \etO\ isotopic abundance ratios in the ISM for different
        evolutionary tracks, predicted using various IMFs. \textbf{a:}
        Star-formation history for a delayed starburst, starting two Gyr after
        the Big Bang, with a total cessation of subsequent star formation.
        Coloured lines correspond to different IMFs. \textbf{b:} Theoretical
        $^{13}$C/$^{18}$O abundance ratio in the ISM as a function of time,
        following the different IMFs (shown in colour). Red symbols refer to
        SMGs in our sample. Blue symbols show the $^{13}$CO/C$^{18}$O ratios
        measured in SMM\,J2135$-$0102 and a few representative local galaxies.
        Table~\ref{tab:imf} lists the detailed definitions of the IMFs adopted
        here. } 
\label{fig:chemical}
\end{figure}

We thus conclude that emission in both $^{13}$CO\ and C$^{18}$O\ is optically
thin for the bulk of the molecular gas mass in these galaxies.  The
systematically low $I$($^{13}$CO)/$I$(C$^{18}$O) ratios found in all our
high-redshift starbursts -- as well as in local ULIRGs -- reflect intrinsic
isotopologue abundance ratios over galaxy-sized molecular hydrogen reservoirs,
i.e.\ $I$($^{13}$CO)/$I$(C$^{18}$O) $\approx$ $^{13}$CO/C$^{18}$O\ $\approx$
$^{13}$C/$^{18}$O. Fig.~\ref{fig:ratios} thus reflects a strong decrease of the
$^{13}$C/$^{18}$O\ abundance ratio in starburst galaxies, compared to local
spiral galaxies, galactic disks and dwarf galaxies.

The only plausible basis for such systematic variations of isotopologue
abundance ratios over galaxy-sized molecular hydrogen reservoirs is a change of
the stellar IMF, which must cause the intrinsic abundances of isotopic elements
to deviate significantly from those found in ordinary star-forming systems
where the standard IMF prevails. The robustness of this conclusion is made
possible by tremendous advances in chemical evolution modeling, which now takes
into account nucleosynthesis, isotopic yields across stellar mass, time
differentials for their release into the ISM, and the dependence of stars'
initial metallicity on prior galactic evolution.  Benchmarked against the rich
isotopic datasets of the Milky Way, these models can now follow the chemical
evolution of various isotopes and their abundance ratios, uniquely identifying
the effects of different IMFs upon them\cite{Romano2017} (see Methods).

In Fig.~\ref{fig:chemical}, we present chemical evolution models that show how
the isotopologue abundance ratios are altered by IMF types, and how they evolve
as a function of cosmic time. The models show a massive galaxy that started an
intense burst of star formation at $z\sim 3$, reached a stellar mass of
$10^{11}$\,M$_{\odot}$\ one Gyr later, ceased forming stars after
the burst, then evolved passively to the present day. This represents an 
extreme case for the evolution of the $^{13}$C/$^{18}$O\ abundance ratio for a
pure starburst. The $^{13}$C/$^{18}$O\ abundance ratio starts from a high value
set by the first generations of metal-poor massive stars, because $^{13}$C\ is
released from its primary and secondary nucleosynthesis channels earlier than $^{18}$O,
which is purely a secondary element (see Methods).  The ratio drops quickly
during the starburst, then slowly increases with time, varying by a factor of
2--$3\times$ depending on the adopted IMF and the time interval. The late
increase of the ratio is due mostly to the slow but continuous release of
$^{13}$C\ from LIMS (see Methods), which keep releasing $^{13}$C\ for a long
time after the star formation -- and, at the same time, the $^{18}$O\ pollution
from massive stars -- has ceased; wiggles in the ratio correspond to the
lifetimes of roughly Solar-mass stars.

It is not possible to reproduce the observed $I$($^{13}$CO)/$I$(C$^{18}$O)
ratio in SMGs with a Kroupa IMF (or similar IMFs; see Table~\ref{tab:imf}),
i.e.\ with an IMF that can reproduce the ratios found in the Milky Way and in
the disks of local spiral galaxies\cite{Kroupa2013}. The top-heavy IMF and the
Ballero\cite{Ballero2007} IMF (which can reproduce the chemical abundances of
stars in the Galactic bulge) under-produce the $I$($^{13}$CO)/$I$(C$^{18}$O) ratios
observed in $z\sim 2$--3 starburst galaxies and local ULIRGs, but they can
reproduce the extremely low $I$($^{13}$CO)/$I$(C$^{18}$O) ratio recently
measured \cite{Sliwa2017} in the central 500-pc region of the starburst ULIRG,
IRAS\,13120$-$5453, which has been until now the lowest value reported in
literature.  Note that the average star-formation event may have a less
top-heavy IMF over galactic scales, or a mix of both top-heavy and canonical
IMFs that produces a galaxy-sized average $^{13}$C/$^{18}$O\ ratio $\le 1$,
which also applies for the resolved studies of IRAS\,13120$-$5453.

A clear trend is shown in Fig.~\ref{fig:chemical}: the more top-heavy the IMF,
the lower the $I$($^{13}$CO)/$I$(C$^{18}$O) ratio, which is also compatible
with the ratios found in local ULIRGs and the exceptionally small
$I$($^{13}$CO)/$I$(C$^{18}$O) ratio found in the centre of IRAS\,13120$-$5453
\cite{Sliwa2017}. This paints a consistent picture in which a top-heavy IMF
operates within both local ULIRGs and the much more numerous, distant starburst
galaxies, where starburst events can quickly enrich the $^{18}$O\ abundance,
pushing the $^{13}$C/$^{18}$O\ ratio to (or below) unity. A canonical IMF can
never produce a $I$($^{13}$CO)/$I$(C$^{18}$O) ratio close to unity, no matter
what type of star-formation history or at what time along a galaxy's
evolutionary track the measurement is made.

\begin{table}[tbh]
\caption{Details of the IMFs used in this work.}
\label{tab:imf}
\begin{center}
\begin{tabular}{l@{\hskip 0.3em}l@{\hskip 0.3em}l@{\hskip 0.3em}l@{\hskip 0.3em}l@{\hskip 0.3em}l@{\hskip 0.3em}l@{\hskip 0.3em}l@{\hskip 0.3em}c}\hline\hline
        IMF Name                    & $\alpha$0 & $\alpha$1 & $\alpha$2 & $m$0  & $m$1  & $m$2  & $m$3 & $M_\star ^{(8-100) M_\odot}$/$M_\star^{\rm total}$ \\
                                    &           &           &           & M$_{\odot}$ & M$_{\odot}$ & M$_{\odot}$ & M$_{\odot}$&   (\%)  \\
\hline
Bottom heavy                        & $-$1.7    & $-$1.7    & $-$1.7    & 0.1   & 0.5   & 1.0   & 100  & 3.9   \\
Kroupa\cite{Kroupa2013,Romano2017}  & $-$0.3    & $-$1.2    & $-$1.7    & 0.1   & 0.5   & 1.0   & 100  & 6.9   \\
Top heavy                           & $-$0.3    & $-$1.1    & $-$1.1    & 0.1   & 0.5   & 1.0   & 100  & 33.3  \\
Ballero\cite{Ballero2007}           & $-$0.3    & $-$0.95   & $-$0.95   & 0.1   & 0.5   & 1.0   & 100  & 44.0  \\
\hline
\end{tabular}\\
\vspace{2em}
\raggedright The slopes quoted in the table are for IMFs in mass, where $m$0 ($=
0.1$ M$_{\odot}$) and $m$3 ($= 100$ M$_{\odot}$) are, respectively, the lower and upper
limits of stellar masses assumed in the models, i.e.\ the IMF is normalised to
unity in the 0.1--100 M$_{\odot}$\ range. $m$1 and $m$2 indicate the masses at
which there is a change in the IMF slope, if any. For instance, for the Kroupa
IMF, the slope changes at both $m$1 and $m$2; for the Ballero and top-heavy
IMFs, the slope changes only at $m$1; finally, the bottom-heavy IMF has a
single slope. The Kroupa IMF slopes are adopted for reproducing typical Milky
Way values in chemical evolution models\cite{Romano2017}, which are within the
error bars of the original reported values\cite{Kroupa2013}.
\end{center}
\end{table}

Multiple evidence in the local Universe has shown that the stellar IMF in
galaxies with very high SFR densities are likely biased to massive stars, such
as ultra-compact dwarf galaxies\cite{Dabringhausen2009},
ULIRGs\cite{Dabringhausen2012}, and progenitors of early-type
galaxies\cite{Peacock2017}. A top-heavy stellar IMF was recently also found in
compact stellar associations in the
LMC\cite{Banerjee2012,Schneider2018,Kalari2018}, whose high-density
star-formation events may closely replicate what happens over galactic scales
in distant starbursts. Our results -- for the most intensive star-forming
systems in the distant Universe, where classical UV and optical methods cannot
be applied -- are in line with these findings.  We also note that metal-poor
dwarf galaxies likely have an IMF biased towards low-mass stars, which is
predicted by the integrated galaxy-wide IMF theory and is consistent with the
results found in dwarf galaxies\cite{Lee2009} and the outer regions of disk
galaxies, using H$\alpha$ and UV observations \cite{PA2008}.

An IMF biased to massive stars implies that SFRs determined for SMGs must be
reduced considerably, since they are based on extrapolations of observables
related to massive stars\cite{Kennicutt1998}. Moving from the Kroupa IMF to the
Ballero IMF, the relative mass fraction of massive stars increases by a factor
of 6--$7\times$ (see Table~\ref{tab:imf}), meaning that SFRs derived from most
classical tracers\cite{Kennicutt1998} (e.g.\ $L_{\rm IR}$, radio continuum,
etc.) must decrease by a similar factor.  As a result, dusty starburst galaxies
probably lie much closer to the so-called `main sequence' of star-forming
galaxies\cite{Noeske2007} than previously thought.  Classical ideas about the
evolutionary tracks of galaxies\cite{Speagle2014} and our understanding of
cosmic star-formation history\cite{Madau1996} are challenged. Fundamental
parameters governing galaxy formation and evolution -- SFRs, stellar masses,
gas-depletion and dust-formation timescales, dust extinction laws, and
more\cite{PA09} -- must be re-addressed, exploiting recent advances in stellar
physics.

\end{bibunit}

\clearpage

\setcounter{table}{0}

\setcounter{figure}{0}

\makeatletter
\renewcommand{\figurename}{Extended Data Fig.}
\renewcommand{\tablename}{Extended Data Table.}

\makeatother

\begin{methods}

\newbibstartnumber{31}
\begin{bibunit}[plainnat]

\section{Sample}

Our sample comprises the strongest CO emitters in the early Universe,
taken from the literature\cite{Magain1988,Weiss2013,Negrello2010}:
four strongly-lensed submillimetre galaxies (SMGs) at $z\sim2$--3,
with look-back times, $\gtrsim 10$\,Gyr. Two of these galaxies,
SPT-S\,J010312$-$4538.8 ($z=3.09$, also known as SPT\,0103$-$45) and
SPT-S\,J012506$-$4723.7 ($z=2.51$, also known as SPT\,0125$-$47), were
selected\cite{Weiss2013} using the South Pole Telescope at
$\lambda = 1.4$ and 2\,mm; another, HATLAS\,J090302.9$-$014127
($z=2.31$, also known as SDP.17b), was discovered using the {\it
  Herschel Space Observatory}\cite{Griffin2010} at far-IR wavelengths;
the last, H1413$+$117 ($z=2.56$, the `Cloverleaf' quasar), was
discovered as a result of its rare quadruple-spot optical morphology,
and was later found to be bright in CO and in dust
continuum\cite{Magain1988,Solomon2003}. We list the basic
characteristics of the sample in Extended Data Table~\ref{tab:targets}.

\begin{table*}[tbh]
\caption{Target properties}\label{tab:targets} 
\begin{center}
        \vspace{5mm}
\fontsize{7pt}{10}\selectfont
\begin{tabular}{llccccccl}\hline
Short name     & IAU name                   & R.A.        & Dec.          & Redshift & Lensing                & $L_{\rm IR}$/$\mu$     & $M_{\star}$/$\mu$      \\  
               &                            & J2000       & J2000         & $z$      & amplification, $\mu$   & $10^{13}$\,L$_{\odot}$ & $10^{10}$\,M$_{\odot}$ \\  
\hline
SPT\,0103$-$45 & SPT-S\,J010312$-$4538.8    & 01:03:11.50 & $-$45:38:53.9 & 3.0917   & $5.3\pm0.11$           & 1.2                    & $5.5^{+6.1}_{-2.9}$    \\  
SPT\,0125$-$47 & SPT-S\,J012506$-$4723.7    & 01:25:07.08 & $-$47:23:56.0 & 2.5148   & $5.5\pm0.1$$\star$     & 2.2                    & --                     \\  
SDP.17b        & HATLAS\,J090302.9$-$014127 & 09:03:03.02 & $-$01:41:26.9 & 2.3051   & $3.56^{+0.19}_{-0.17}$ & 2.0                    & $24.2^{+8.6}_{-4.0}$   \\  
Cloverleaf     & H1413$+$117                & 14:15:46.23 & $+$11:29:44.0 & 2.5585   & 11$\dagger$            & 6.0                    & --                     \\  
\hline
\end{tabular}\\
\end{center}
\noindent Data for SDP.17b is from ref. \cite{Dye2014,Negrello2014}. \\
\noindent$\star$ We adopt an amplification factor, $\mu$, derived from lens modelling using 850-$\mu$m ALMA
visibility data in the literature\cite{Ma2015,Aravena2016}. \\
\noindent$\dagger$ We adopt the best-determined amplification factor, $\mu$,
reported in the literature\cite{VS2003,Solomon2003}, whose lensing model is
derived from CO $J=7\rightarrow 6$ line emission, and could better reproduce a single
source on the lens plane\cite{VS2003}. Unfortunately the uncertainty of this
amplification factor was not reported, but the uncertainty in $\mu$ does not
jeopardise our conclusions.
\end{table*}

\section{Observations and data reduction}

We have performed simultaneous observations of $^{13}$CO\ and C$^{18}$O\ using
ALMA in its relatively compact array configurations (C36-1 and C36-2), with two
2-GHz-wide spectral windows (SPWs) in bands 3 and 4.  We used the remaining two
SPWs to cover continuum emission. Between 10 and 30\,min were spent on target
for each transition. For SDP.17b, we observed both $J=3\rightarrow 2$ and
$J=4\rightarrow 3$ transitions of $^{13}$CO\ and C$^{18}$O, in order to have
a redundant measurement for their line ratios as well as constraints on the
relative excitation in these rare isotopic lines. We also observed $^{12}$CO\
$J=5\rightarrow 4$ for SPT\,0103$-$45 and $^{12}$CO\ $J=4\rightarrow 3$
for SPT\,0125$-$47, with similar configurations.  Calibrators, integration
time, atmospheric conditions are listed in Extended Data
Table~\ref{tab:observations}.

All the data were calibrated manually using CASA v4.7.1 \cite{McMullin2007},
using standard procedures. We subtracted the continuum using the CASA task,
{\sc uvcontsub}, by fitting a linear slope to the line-free channels. We
cleaned the visibility data with a channel width of $\sim20$--30\,km\,s$^{-1}$,
using a Briggs weighting with {\sc robust} = 1.5 to optimise sensitivity. We
applied a primary beam correction to all the cleaned data. Our target galaxies
are mostly unresolved, or only marginally resolved.  We assume that line widths
of $^{13}$CO\ and C$^{18}$O\ are the same as those of $^{12}$CO\ lines, to
minimise uncertainties in the line flux fitting.  Extended Data
Figs~\ref{fig:sdp17b}, \ref{fig:spt0103} and \ref{fig:cloverleaf} present the
velocity-integrated flux (moment-0) maps of $^{13}$CO\ and C$^{18}$O, overlaid
with contours of high-resolution submm continuum.  In Extended Data
Fig.~\ref{fig:spectra} we present the spectra.  SPT\,0103$-$45 has two velocity
components that cover a very large velocity span. The overall line profile of
$^{13}$CO\ is consistent with $^{12}$CO, but limited by the noise level. We
adopt only the narrow (stronger) component, seen for the yellow shadow region
in the $^{12}$CO\ $J=5\rightarrow 4$ spectrum, to avoid confusion from the broad
(weaker) component (see Extended Data Fig.~\ref{fig:spectra} ). Our synthesised
beamsizes are mostly larger than, or at least comparable to, the apparent sizes
revealed by the high-resolution submm continuum images, so any missing flux is
expected to be negligible.  We extract spectra of $^{13}$CO\ and C$^{18}$O\
using circular apertures $\approx 4$--6$''$ in diameter, as shown in Extended Data
Figs~\ref{fig:sdp17b}, \ref{fig:spt0103} and \ref{fig:cloverleaf}.

To measure velocity-integrated line fluxes and the associated errors, we
performed three independent methods: we first fit one-dimensional Gaussian
profiles to the extracted spectra with a fixed linewidth from $^{12}$CO, and
fixed the frequency interval between $^{13}$CO and C$^{18}$O, since we can assume
confidently that these two lines are emitted from the same excitation
component, such that their line centres do not shift relative to one another.
Second, we made moment-0 maps and fitted two-dimensional Gaussian profiles, as an
independent check of line flux. Third, to better estimate the noise level, we
also calculated the theoretical noise using the ALMA sensitivity
calculator\cite{ALMAcalculator}, given the integration time, precipitable water
vapour, linewidth, and array configuration. The measured line fluxes and
properties are listed in Extended Data Table~\ref{tab:fluxes}.  To be
conservative, in Figs~\ref{fig:ratios} and \ref{fig:chemical} we adopt the
largest error among the results from the three methods in the analysis.

\section{Line ratios and optical depths} 

\subsection{Conditions of local thermodynamic equilibrium (LTE)} 

We first analyse the observed line ratios of $^{13}$CO\ to C$^{18}$O, to
constrain the molecular line optical depths, assuming local thermodynamic
equilibrium (LTE; $T_{\rm ex}^{\rm line} = T_{\rm kin}^{\rm line}$). The line
brightness temperature ratios of $^{12}$CO\ to $^{13}$CO\ and $^{13}$CO\ to
C$^{18}$O can be expressed as:

\begin{equation}
\frac{T_{\rm b}^{\rm ^{12}CO}}
     {T_{\rm b}^{\rm ^{13}CO}} 
= 
 \frac{J_\nu(T_{\rm ex} ^{\rm ^{12}CO}) - J_\nu(T_{\rm bg} ^{\rm ^{12}CO})}
      {J_\nu(T_{\rm ex} ^{\rm ^{13}CO}) - J_\nu(T_{\rm bg} ^{\rm ^{13}CO})} 
\cdot
\frac{1-\exp (-\tau ^{\rm ^{12}CO})} 
     {1-\exp (-\tau ^{\rm ^{13}CO})}
\end{equation}

and 

\begin{equation}
\frac{T_{\rm b}^{\rm ^{13}CO}}
 {T_{\rm b}^{\rm C^{18}O}} 
= 
\frac{J_\nu(T_{\rm ex} ^{\rm ^{13}CO}) - J_\nu(T_{\rm bg} ^{\rm ^{13}CO})}
 {J_\nu(T_{\rm ex} ^{\rm C^{18}O}) - J_\nu(T_{\rm bg} ^{\rm C^{18}O})} 
\cdot
\frac{1-\exp (-\tau ^{\rm ^{13}CO})} 
 {1-\exp (-\tau ^{\rm C^{18}O})}
\end{equation}

\noindent where $T_{\rm b}$ is the brightness temperature of the molecular transition,
$T_{\rm ex}$\ is the excitation temperature, $J$ is the rotational quantum number, and
$T_{\rm bg}$\ is the radiation temperature of the background emission field, which is
dominated by the cosmic microwave background (CMB) at $T_{\rm CMB}$\ $\sim 2.73 \times
(1+z)$ K, where $z$ is the redshift. $\tau^{\rm line}$ is the optical depth of the given transition.
$J(T)=(h\nu/k_{\rm B})/[\exp(h\nu/k_{\rm B}T)-1]$ is the Planck radiation
temperature at the rest frequency of the line emission, $\nu^{\rm line}$.
$k_{\rm B}$ is the Boltzmann constant, $h$ is the Planck constant, and $T$ is
the considered temperature. For optically thick lines (e.g.\ $^{12}$CO for most
conditions), $1-\exp (-\tau ^{\rm line}) \sim 1$; for optically thin lines
(e.g.\ $^{13}$CO\ and C$^{18}$O), $1-\exp (-\tau ^{\rm ^{line}}) \sim \tau^{\rm
line}$.

In Extended Data Fig.~\ref{fig:LTE} we present the
$I$($^{13}$CO)/$I$(C$^{18}$O)\ and $I$($^{12}$CO)/$I$($^{13}$CO)\
velocity-integrated line intensity ratios as a function of the optical depths
of $^{13}$CO\ and C$^{18}$O, under LTE conditions. We also calculated the
corresponding H$_2$\ column densities, assuming Galactic
abundances\cite{Mangum2015,Frerking1982}. A representative Galactic abundance
ratio of $^{12}$CO/$^{13}$CO $=70$ is assumed, the optical depth for $^{13}$CO\
needs to be $<0.03$, which is around $10\times$ lower than the Galactic average
values\cite{Barnes2015} for producing the observed high
$I$($^{12}$CO)/$I$($^{13}$CO) ratios, $\ge 30$. The corresponding optical depth
of $^{12}$CO\ is $\sim 2$, much lower than the typical values for $^{12}$CO\
$J=1\rightarrow 0$ found in typical Galactic molecular clouds\cite{Barnes2015},
but consistent with a moderate optical depth of $^{12}$CO\ found in local
starburst galaxies\cite{Aalto1995}.

Only when the optical depth of C$^{18}$O\ $\gg 1$ (corresponding to $\tau^{\rm
^{13}CO}$\ $\gg 7$, which leads to an H$_2$\ column density of $\gg 10^{25}$
cm$^{-2}$), does the $I$($^{13}$CO)/$I$(C$^{18}$O) line ratio approach unity.
In this case the line ratios of $I$($^{12}$CO)/$I$($^{13}$CO) and
$I$($^{12}$CO)/$I$(C$^{18}$O) would also move towards unity, in conflict with
our observed ratios. Even for moderate $\tau^{\rm ^{13}CO}\sim 0.2$--0.5, the
line ratio of $I$($^{13}$CO)/$I$(C$^{18}$O) stays at $\sim 6$--7, not strongly
biased by $\tau^{\rm ^{13}CO}$.

\subsection{Non-LTE conditions}

In Extended Data Fig.~\ref{fig:tau}, we present non-LTE models derived with a
non-LTE radiative transfer code, RADEX\cite{vdTak2007}, showing the optical
depth, $I$($^{13}$CO)/$I$(C$^{18}$O) and $I$($^{12}$CO)/$I$($^{13}$CO) line
ratios as a function of $^{13}$CO\ column density and $N_{\rm H_2}$\ column
density, $N_{\rm H_2}$. We calculate different H$_2$\ volume densities, $n_{\rm H_2}$, of
10$^3$ cm$^{-3}$, 10$^4$ cm$^{-3}$ and 10$^5$ cm$^{-3}$, covering the most common $n_{\rm H_2}$\
range -- typical values from normal molecular clouds to dense cores. We assume
that the same abundance ratios as we assumed for the LTE conditions, i.e.
$^{12}$CO/$^{13}$CO\ = 70, and $^{13}$CO/C$^{18}$O\ = 7, which are
representative values of the Milky Way disk. The velocity width (full-width at
half-maximum, {\sc fwhm}) is set to 300 km\,s$^{-1}$, as the typical (indeed,
at the lower end) linewidth found in ULIRGs and SMGs\cite{Yang2017}. In the
panels \textbf{b} and \textbf{c} of Extended Data Fig.~\ref{fig:tau} we overlay
the LTE results, for comparison.

For all models, we set the kinetic temperature, $T_{\rm kin}$, to be 30\,K,
which is a typical dust temperature for the SMG population\cite{Simpson2017},
and is also the lower limit of the kinetic temperate of the H$_2$\ gas, as the
minimum temperature powered by the cosmic-ray heating for such starburst
conditions\cite{PPP2014}. Higher $T_{\rm kin}$\ would bring the CO energy
population towards higher-$J$ transitions, making optical depths even smaller.

Extended Data Fig.~\ref{fig:tau} shows that, for $n_{\rm H_2} = 10^3$
cm$^{-3}$, which is a typical value for normal Galactic molecular cloud
conditions, only when the H$_2$\ column density $N_{\rm H_2} \gg 10^{26}$
cm$^{-2}$, the line ratio of $I$($^{13}$CO)/$I$(C$^{18}$O) can approach unity
($< 1.5$, considering uncertainties in line ratios). The required high column
densities are a few orders of magnitude higher than the typical values measured
in SMGs: $\sim 10^{23}$--$10^{24}$ cm$^{-2}$, obtained using
X-rays\cite{Wang2013}, CO LVG modeling\cite{Spilker2014}, and
dust\cite{Simpson2017}. This is especially supported by the Cloverleaf quasar,
whose X-ray emission has been clearly detected \cite{Chartas2004}, given the
Compton limit of $\sim 10^{24}$ cm$^{-2}$.

On the other hand, the high-density results, i.e.\ $n_{\rm H_2} =10^4$ and $10^5$
cm$^{-3}$, are very similar to those under LTE conditions.  For all conditions,
a moderate $^{13}$CO\ optical depth does not strongly vary the
$I$($^{13}$CO)/$I$(C$^{18}$O) ratio from the abundance ratio. So, it is highly
unlikely that the unity value of the $I$($^{13}$CO)/$I$(C$^{18}$O) ratio can be
due to high optical depths.

\section{Possible HNCO contamination of C$^{18}$O\ lines}

HNCO $5_{0 5}$ $\rightarrow$\ $4_{04}$ ($J= 5$$\rightarrow$\,4) has a rest
frequency of 109.9058\,GHz, close to the rest frequency of C$^{18}$O\
$J=1\rightarrow 0$ (109.7822 GHz), with a velocity offset of $\sim 370$
km\,s$^{-1}$. When the linewidths are broad, these two lines are sometimes
blended, which leads to a possible contamination of the C$^{18}$O\
measurements\cite{Sliwa2017,Martin2009,Greve2009}.  In these observations it
has been found that HNCO $J= 5\rightarrow 4$ could contribute up to $\sim
30$\% of the total flux (C$^{18}$O\ + HNCO), for the most extreme cases in the
local Universe, e.g.\ Arp\,220\cite{Greve2009} and
IRAS\,13120$-$543\cite{Sliwa2017}.

HNCO $J= 5\rightarrow 4$ has a critical density, $n_{\rm crit}\sim 10^6$ cm$^{-3}$\
and is regarded as a dense-gas tracer that could be excited in slow-shock
regions\citep{Zinchenko2000,Li2013}.  The Einstein $A$ coefficient increases
as $A \propto  (J+1)^3 $, thus $n_{\rm crit}$ increases quickly for high-$J$
transitions, such as HNCO $J=15\rightarrow 14$ ($\nu_{\rm rest} = 329.66$
GHz) and $J=20\rightarrow 19$ ($\nu_{\rm rest} = 439.62$ GHz), making much
less contribution to the C$^{18}$O\ $J=3\rightarrow 2$ and
$J=4\rightarrow 3$ transitions involved in our study.

To better estimate how much HNCO lines may contaminate the C$^{18}$O lines, in
Extended Data Fig.~\ref{fig:hnco} we show the theoretical line ratios between
HNCO and C$^{18}$O\ and high-$J$ $I$(HNCO)/$I$(C$^{18}$O) ratios normalised
with $I$(HNCO $J=5\rightarrow 4$)/$I$(C$^{18}$O\ $J=1\rightarrow 0$), using
RADEX\cite{vdTak2007}. We assume the same abundances measured in
Arp\,220\cite{Martin2009}, and use molecular data from the Leiden Atomic and
Molecular Database\cite{Schoier2005} (LAMBDA).  We assume $T_{\rm kin}$\ =
30\,K as the representative temperature of the H$_2$\ gas.
 
Extended Data Fig. \ref{fig:hnco}\textbf{a} shows that the
$I$(HNCO)/$I$(C$^{18}$O) ratio increases with the volume density of H$_2$ gas,
$n_{\rm H_2}$. Moreover, the ratio decreases quickly with $J$ transitions,
meaning that the contamination from HNCO to C$^{18}$O\ is much less severe for
the high-$J$ transitions, compared to the C$^{18}$O\ $J=1\rightarrow 0$ line.
Extended Data Fig. \ref{fig:hnco}\textbf{b} shows $I$(HNCO)/$I$(C$^{18}$O) line
brightness ratios normalised by $I$(HNCO $J=5\rightarrow 4$)/$I$(C$^{18}$O\
$J=1\rightarrow 0$). With a weak dependency on $n_{\rm H_2}$,  the ratios of
$I$(HNCO $J=10\rightarrow 9$)/$I$(HNCO $J=5\rightarrow 4$) and $I$(HNCO
$J=15\rightarrow 14$)/$I$(HNCO $J=5\rightarrow 4$) are one order of magnitude
lower than unity. Unfortunately the LAMBDA database does not have the data for
the transition of HNCO $J=20\rightarrow 19$, whose HNCO/C$^{18}$O\, ratio is
expected to be even lower.

Even if we take the highest HNCO $J=5\rightarrow 4$ contamination found in
local galaxies, i.e.\ 30\%, the corresponding contamination for the high-$J$
C$^{18}$O transitions will be at most 3\%, which can be regarded as negligible
for the lines in our study.

\section{Chemical evolution model}

We adopt a single-zone chemical evolution model for our analysis,
originally developed to describe the evolution of the Milky
Way\cite{Matteucci2012}, then further extended to other
galaxies\cite{Romano2015}.  The model computes the evolution of
abundances of multiple elements, including $^{12}$C, $^{16}$O,
$^{13}$C\ and $^{18}$O\ in the ISM of galaxies. We use detailed
numerical models to solve the classical set of equations of chemical
evolution\cite{Tinsley1980,Pagel1997,Pagel2009,Matteucci2001,Matteucci2012,Romano2015},
with the following assumptions:

\begin{itemize} 
\item Gas inflow with primordial chemical composition provides raw material for
        star formation. The gas is accreted at an exponentially fading rate and
        the timescale of the process is a free parameter of the model;  
\item Galactic outflows remove both the stellar ejecta and a fraction of the
        ambient ISM; 
\item Star formation follows the canonical Kennicutt-Schmidt
        law\cite{Kennicutt1998b}; the masses of the newly-formed stars follow
        the input IMF; 
\item Finite stellar lifetimes for different stars need to be considered (i.e.\
        no instantaneous recycling approximation (non-IRA) is adopted)\cite{Schaller1992}; 
\item Stars release the elements they have synthesised during their lifetime,
        as well as those already present when they were born that are left
        unaltered by the nucleosynthesis processes, when they die; 
\item Stellar ejecta are mixed with the ISM homogeneously.  
\end{itemize} 

The adopted yields account for the dependence of several stellar
processes on the initial metallicity of the stars, and have been
calibrated with the best fit using the Milky Way data, which are
relevant to a range of metallicity and evolution
timescales\cite{Romano2017}.  The time-delay effect is considered in
the chemical evolution, namely the differences between the lifetimes
of massive stars and low-mass stars\cite{Matteucci1986}. We used an
analytical formula for the stellar lifetimes that linearly interpolate
stellar lifetime tables\cite{Schaller1992}.  The time lag in producing
and releasing primary (those synthesised directly from H and He; i.e.\
$^{12}$C, $^{16}$O) and secondary elements (those derived from metals
already present in the star at birth; i.e.\ $^{13}$C, $^{18}$O\ -- but
note that a fraction of $^{13}$C\ is also synthesised as a primary
element) is also considered\cite{Matteucci2012}. These two effects,
corroborated by the star-formation history and the adopted IMF,
determine the amount of different isotopes released to the ISM on
different timescales. In particular, different isotopes are released
as a function of time, e.g.\ the bulk of $^{13}$C\ is released later
than $^{12}$C, and the bulk of $^{18}$O is released later than
$^{16}$O\cite{Henkel1993,Wilson1992, Romano2017}. Chemical evolution
models can now follow the evolution of various isotopic ratios,
tracing abundance ratios not only between the isotopes of each
element\cite{Romano2017}, but also between different elements.

The most important aspect regarding this work is that such models can now
compare the effects of a young starburst (with a regular stellar IMF, e.g.\ the
Kroupa IMF) against those due to different stellar IMFs via carefully chosen
isotope (and thus isotopologue) ratios. This critical advance was made by no
longer assuming instantaneous element enrichment of the ISM by the stars, but
incorporating the different timescales of their release into the ISM. It should
be noted that these timescales, and the relative delays between the release of
various isotopes into the ISM, are set by stellar physics, i.e.\ they are not
free parameters. With the Kroupa IMF, only an unphysical combination of
star-formation history, i.e.\ $\tau \lesssim 10$ Myr with an SFR
$\gtrsim 20\,000$\, M$_{\odot}$ yr$^{-1}$ could approach the observed
$^{13}$C/$^{18}$O\, ratios of near unity.

\section{Origins of carbon and oxygen isotopes}

The $I(^{12}$CO)/$I(^{13}$CO) line ratios have been found to vary
systematically in galaxies with different SFRs and Hubble
types\cite{Aalto1995,Davis2014}. Owning to the differences in the origins of
$^{12}$C and $^{13}$C, it has been proposed that the $I(^{12}$CO)/$I(^{13}$CO)
line ratio can be used to derive their abundance ratio, which can further probe
the stellar IMF, or different star-formation modes\cite{Henkel1993,
Henkel2010,PPP2014, Sliwa2017}.

The $^{12}$C element is primarily produced by helium burning (the classical
triple-$\alpha$ process), and multiple channels can produce $^{12}$C in
nucleosynthesis\cite{Hughes2008}. In the Milky Way, $^{12}$C is primarily
produced by LIMSs, revealed by data for stars in the Solar vicinity: in fact,
the [C/Fe]/[Fe/H] ratio is almost a constant, with [C/Fe] being Solar,
indicating that C and Fe are produced in the same proportions by the same
stars\cite{Nomoto2006,Romano2010}. If mass loss from massive stars is
considered, $^{12}$C\ released from massive stars still accounts for $< 50$\%
of the total\cite{Cescutti2009,Carigi2005}. On the other hand, $^{13}$C is
released from LIMSs largely as a secondary element, because $^{13}$C production
needs a pre-existing seed, namely, the primary element, $^{12}$C
\cite{Hughes2008,Meyer2008}.  $^{13}$C also has a primary component in
nucleosynthesis but it can only occur in red AGB stars, where periodic
dredge-up episodes convect $^{12}$C\ to the stellar surface and form $^{13}$C.

Since both $^{12}$C and $^{13}$C are mostly produced by LIMSs, the
$^{12}$C/$^{13}$C ratio cannot discriminate between IMFs unambiguously.
In previous work, however, we show that by switching from the Ballero
IMF\cite{Ballero2007} (a very top-heavy IMF, which can reproduce the chemical
abundances of stars in the Galactic bulge) to the Kroupa IMF, the
$^{12}$C/$^{13}$C ratio only varies\cite{Romano2017} by a factor of $2\times$,
indicating that carbon isotopologue ratios are not very sensitive to IMF.
Furthermore, the $^{12}$C abundance is very difficult to obtain because the
$^{12}$C-bearing major isotopologue lines are mostly optically thick.

The origin of oxygen isotopes is rather different. As earlier work
suggested\cite{Sage1991}, the stellar yields of $^{16}$O and $^{18}$O
are sensitive to different stellar masses, due to their temperature sensitivity
in the stellar nucleosynthesis\cite{Kobayashi2011}. Production of $^{16}$O is
dominated by massive stars\cite{Tinsley1980}, as revealed by chemical evolution
models in the Milky Way using detailed stellar yields\cite{Romano2017}.  Only a
tiny fraction of $^{16}$O is contributed by AGB stars\cite{Meyer2008}.  Massive
stars also dominate the production of $^{18}$O\cite{Timmes1995}, predominantly
in the early stages of helium burning \cite{Kobayashi2011}. As a secondary
element, the $^{18}$O yield relies strongly on the pre-existence of $^{16}$O,
so the metallicity in oxygen also plays a major role in producing $^{18}$O. The
production of $^{18}$O is biased to more massive stars compared to $^{13}$C
which is more biased to LIMSs. So, the abundance ratio of $^{13}$C\ and
$^{18}$O\ does indeed reflect different IMFs (see Fig.~\ref{fig:chemical}).

The abundance ratios of $^{12}$C/$^{13}$C\ and $^{16}$O/$^{18}$O\ can trace
star-formation timing and IMF\cite{Romano2017}, respectively.  The $^{12}$C\
and $^{16}$O\ abundances are compromised by the optical depths of molecular
lines, which are difficult to estimate accurately. However, the combination of
both carbon and oxygen isotopologues -- the abundance ratio of $^{13}$C\ and
$^{18}$O\ -- can be obtained easily from the ratio of two optically thin lines,
$^{13}$CO\ and C$^{18}$O, in the same $J$ transition.  Moreover, these two
lines can be obtained simultaneously using current facilities, due to the close
spacing of their rest frequencies. They have almost identical critical densities and
upper energy levels, essentially free from excitation differences. Even for strongly
lensed galaxies, it is safe to assume that a differential lensing effect among
the two lines is negligible.

\end{bibunit}

\end{methods}

\begin{extendeddata}

\begin{figure*}
\centering
\includegraphics[scale=1.0]{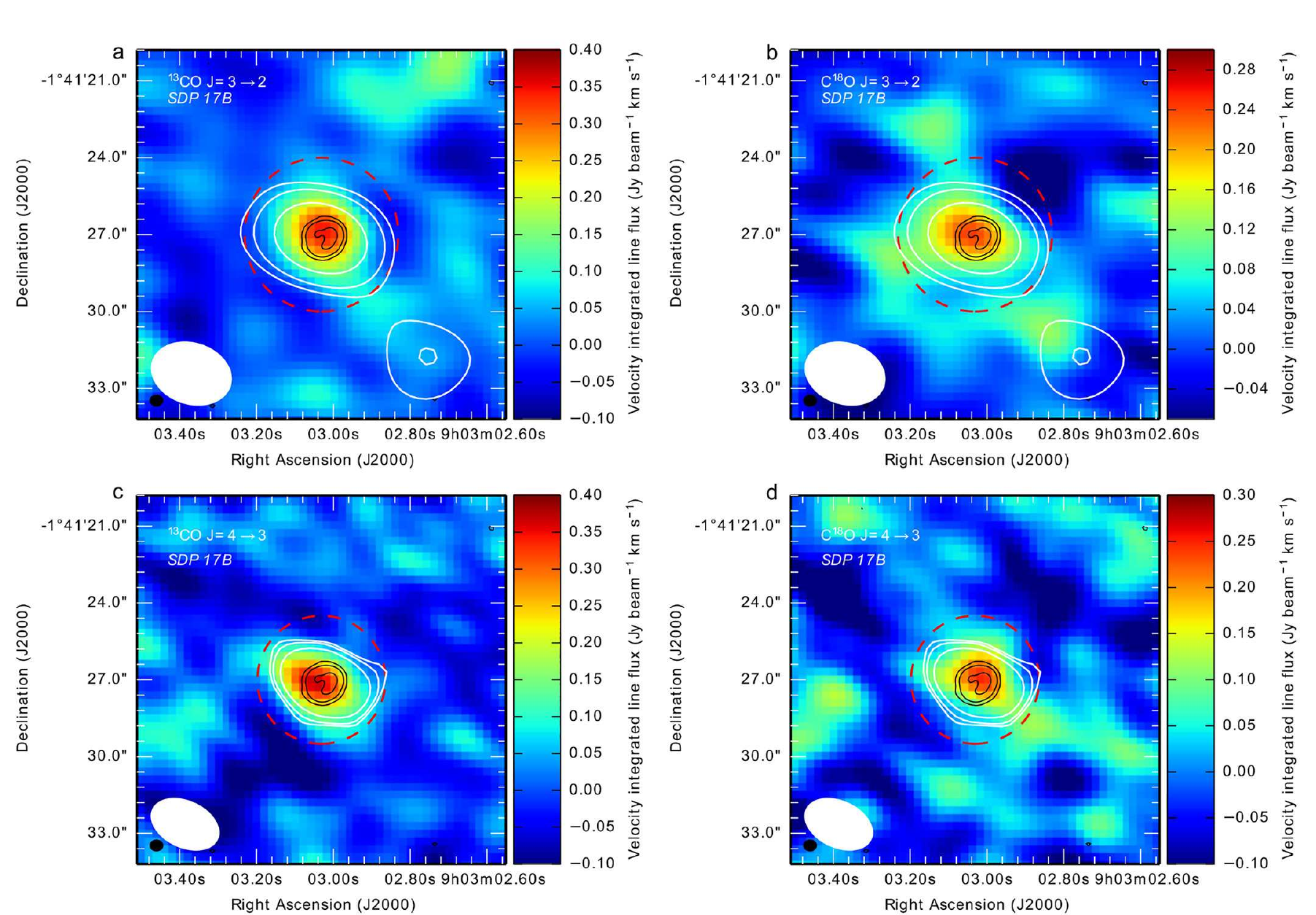}
\caption{Velocity-integrated flux maps (Moment-zero) of 
        SDP~17b.  Black contours show the high resolution 250-GHz continuum
        image, obtained from the ALMA archive 
        \cite{Falgarone2017}, with levels of 3, 10 and 50
        $\sigma$ ($\sigma= 0.6\times 10^{-1}$ mJy beam$^{-1}$). Dashed red
circles show the adopted apertures for extracting spectra.
\textbf{a \& b:}
Images of $^{13}$CO\ and C$^{18}$O\ for the $J=3$$\rightarrow$\,2 transition.  White contours
show the 95-GHz continuum, with levels of 3, 5 and 10 $\sigma$ ($\sigma =
1.7\times 10^{-2}$ mJy beam$^{-1}$).
\textbf{c \& d:}
Images of $^{13}$CO\ and C$^{18}$O\ for the $J=4$$\rightarrow$\,3 transition.  White contours
show the 133-GHz continuum, with levels of 3, 5 and 10 $\sigma$ ($\sigma =
2.3\times 10^{-2}$ mJy beam$^{-1}$).
} \label{fig:sdp17b}
\end{figure*}

\begin{figure*}
\centering
\includegraphics[scale=1.0]{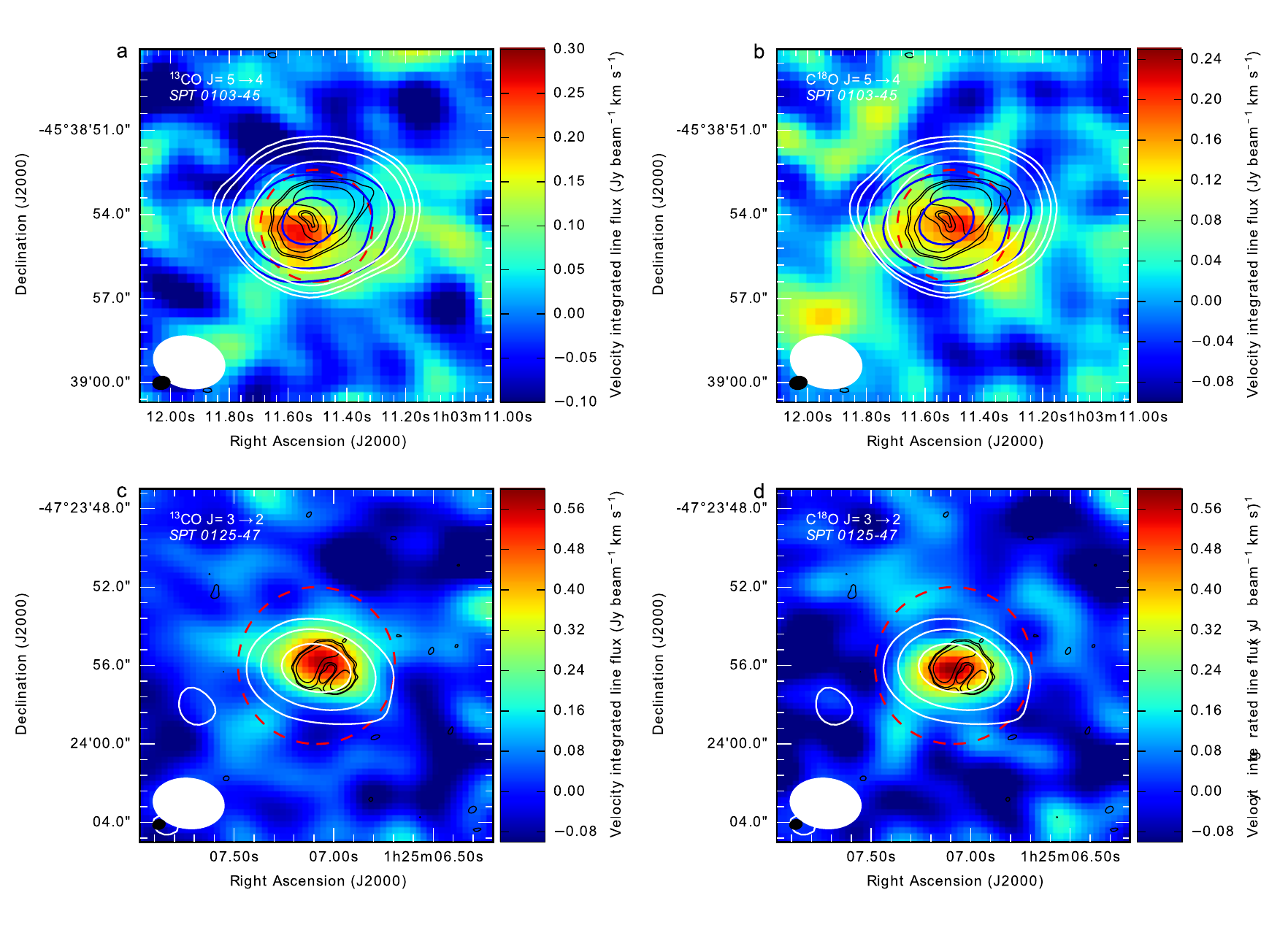}
\caption{
Velocity-integrated flux maps (Moment-zero) of SPT\,0103$-$45 and SPT\,0125$-$47. 
Black contours show the high-resolution 336-GHz continuum image, obtained from
the ALMA archive\cite{Vieira2013}, with levels of 3, 10 and 30 $\sigma$
($\sigma= 2.3\times 10^{-2}$ mJy beam$^{-1}$). Dashed red circles show the
adopted apertures for extracting spectra. 
\textbf{a \& b:} Images of $^{13}$CO\ and C$^{18}$O\ $J=$5$\rightarrow$\,4 for SPT\,0103$-$45.  
Blue contours show the narrow $^{12}$CO\ $J=4$$\rightarrow$\,3 emission, with levels of 3, 10
and 30 $\sigma$ ($\sigma = 0.14$ Jy\, beam$^{-1}$\, km\,s$^{-1}$). 
White contours show the 135-GHz continuum, with
levels of 3, 10 and 30 $\sigma$ ($\sigma =
2 \times 10^{-2}$ mJy beam$^{-1}$).
\textbf{c \& d:} Images of $^{13}$CO\  and C$^{18}$O\ for the $J=3$$\rightarrow$\,2 transition in
SPT\,0125$-$47.  White contours show the 94-GHz continuum, with levels of 3, 5
and 10 $\sigma$ ($\sigma = 2.2\times 10^{-2}$ mJy beam$^{-1}$).} 
\label{fig:spt0103}
\end{figure*}

\begin{figure*}
\centering
\includegraphics[scale=1.0]{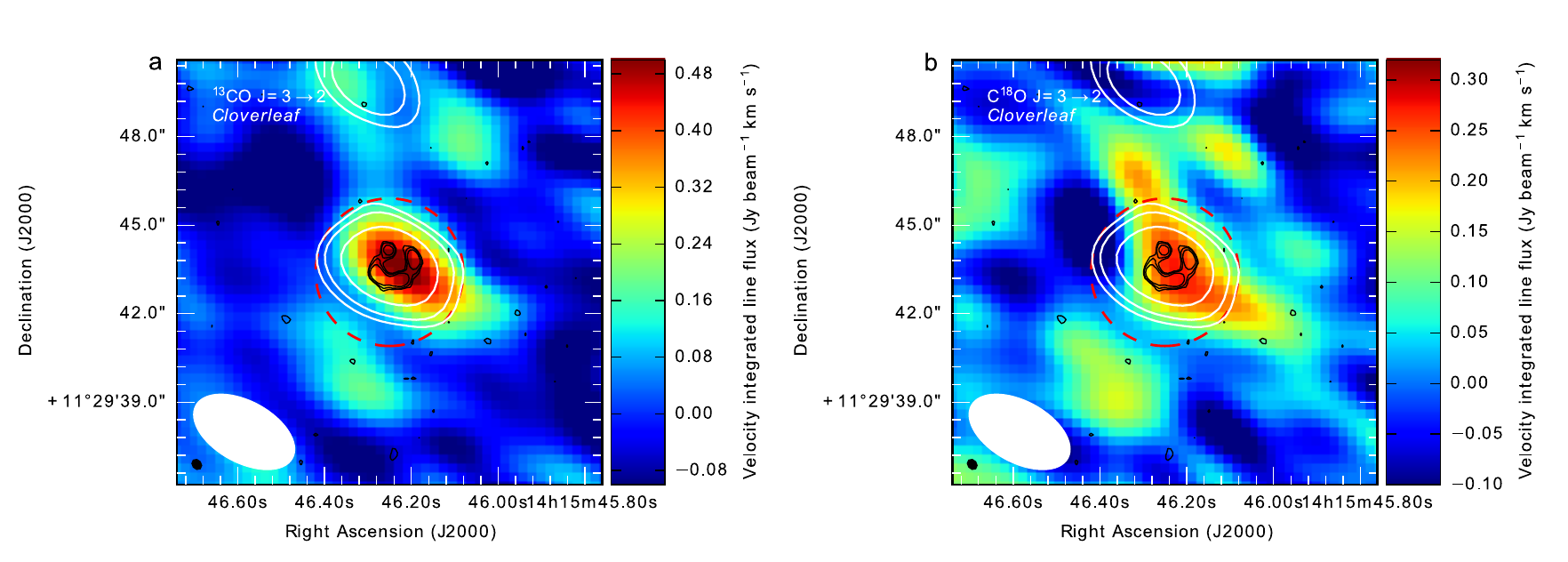}
\caption{
Velocity-integrated flux maps (Moment-zero) of Cloverleaf. 
\textbf{a:}
Image of the $^{13}$CO\ $J=$3$\rightarrow$\,2 transition. 
\textbf{b:}
Image of the C$^{18}$O\ $J=$3$\rightarrow$\,2 transition. 
Black contours show
the high-resolution 690-GHz continuum image, obtained from the ALMA archive
\cite{Ferkinhoff2015}, with levels of 3, 5 and 10 $\sigma$ ($\sigma$= 0.8 mJy
beam$^{-1}$). Dashed red circles show the adopted apertures for extracting
spectra.  White contours show the 92-GHz continuum, with levels of 3, 5 and 10
$\sigma$ ($\sigma = 2\times 10^{-2}$ mJy beam$^{-1}$).
} \label{fig:cloverleaf}
\end{figure*}

\begin{figure}
\centering
\includegraphics[scale=0.95]{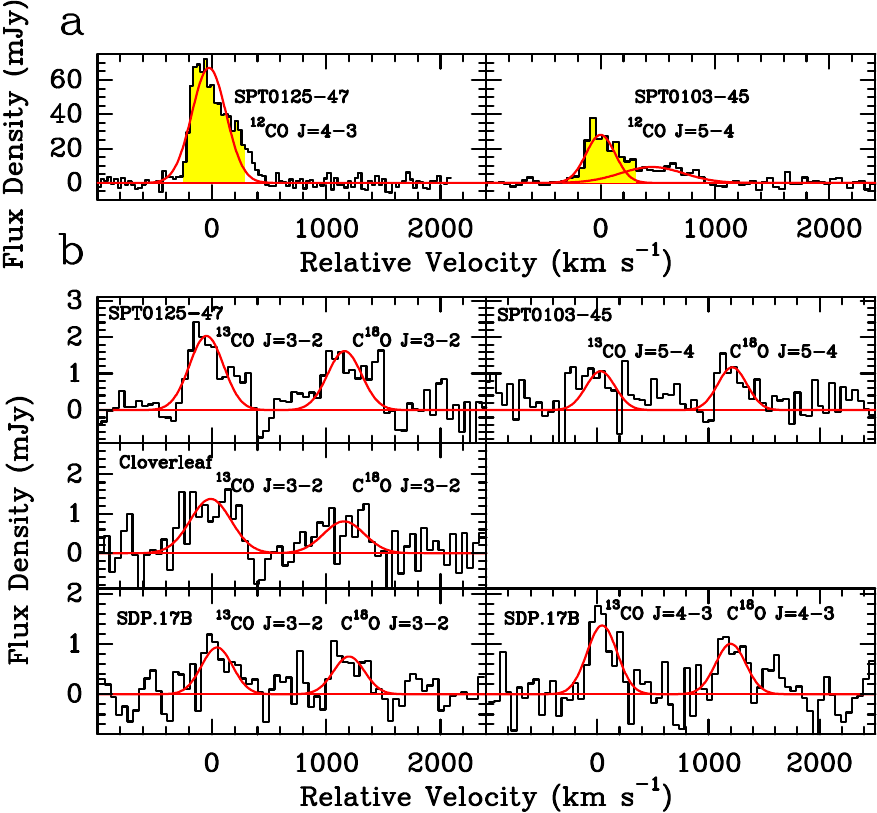}
 \caption{\textbf{a:} ALMA spectra of $^{12}$CO\ in SPT\,0125$-$47 and SPT\,0103$-$45.
         Yellow shadows show the velocity range adopted from $^{12}$CO\ in the analysis. 
         \textbf{b:} ALMA spectra of $^{13}$CO\ and C$^{18}$O\ for all targets. All
         spectra are in black. Red lines show Gaussian fits to the observed
         lines. Velocities are labelled relative to their $^{12}$CO\ or $^{13}$CO\
         transitions.}
 \label{fig:spectra}
\end{figure}

\begin{figure}
\centering
\includegraphics[scale=0.95]{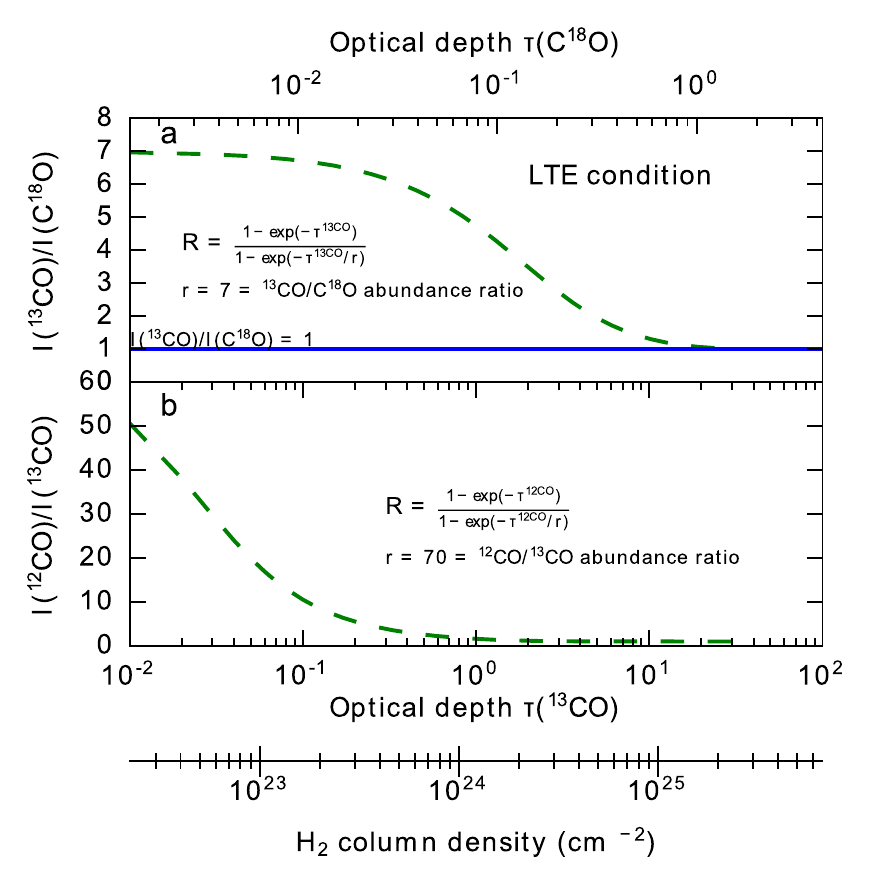}
\caption{ 
\textbf{a:} $I$($^{13}$CO)/$I$(C$^{18}$O) line ratio as a function of optical
depth of $^{13}$CO. \textbf{b:}  $I$($^{12}$CO)/$I$($^{13}$CO)  line ratio as a
function of optical depth of $^{13}$CO. Both ratios assume local thermal
equilibrium conditions.  We assume abundance ratios of $^{13}$CO/C$^{18}$O\ and
$^{12}$CO/$^{13}$CO\ are 7 and 70, respectively,  which are representative
values found in the Milky Way.  This shows that the
$I$($^{13}$CO)/$I$(C$^{18}$O)\ line ratio approaches unity (blue line) only
when the optical depth of C$^{18}$O\ is $\ge$ 1 (and the corresponding
$\tau^{\rm ^{13}CO}$\ = 7.). The bottom scale bar shows the corresponding
column density of H$_2$\ gas, assuming a CO/H$_2$\ abundance of 8.5
$\times10^{-5}$ \cite{Frerking1982}.} \label{fig:LTE}
\end{figure}

\begin{figure*}
\centering
\includegraphics[scale=1.0]{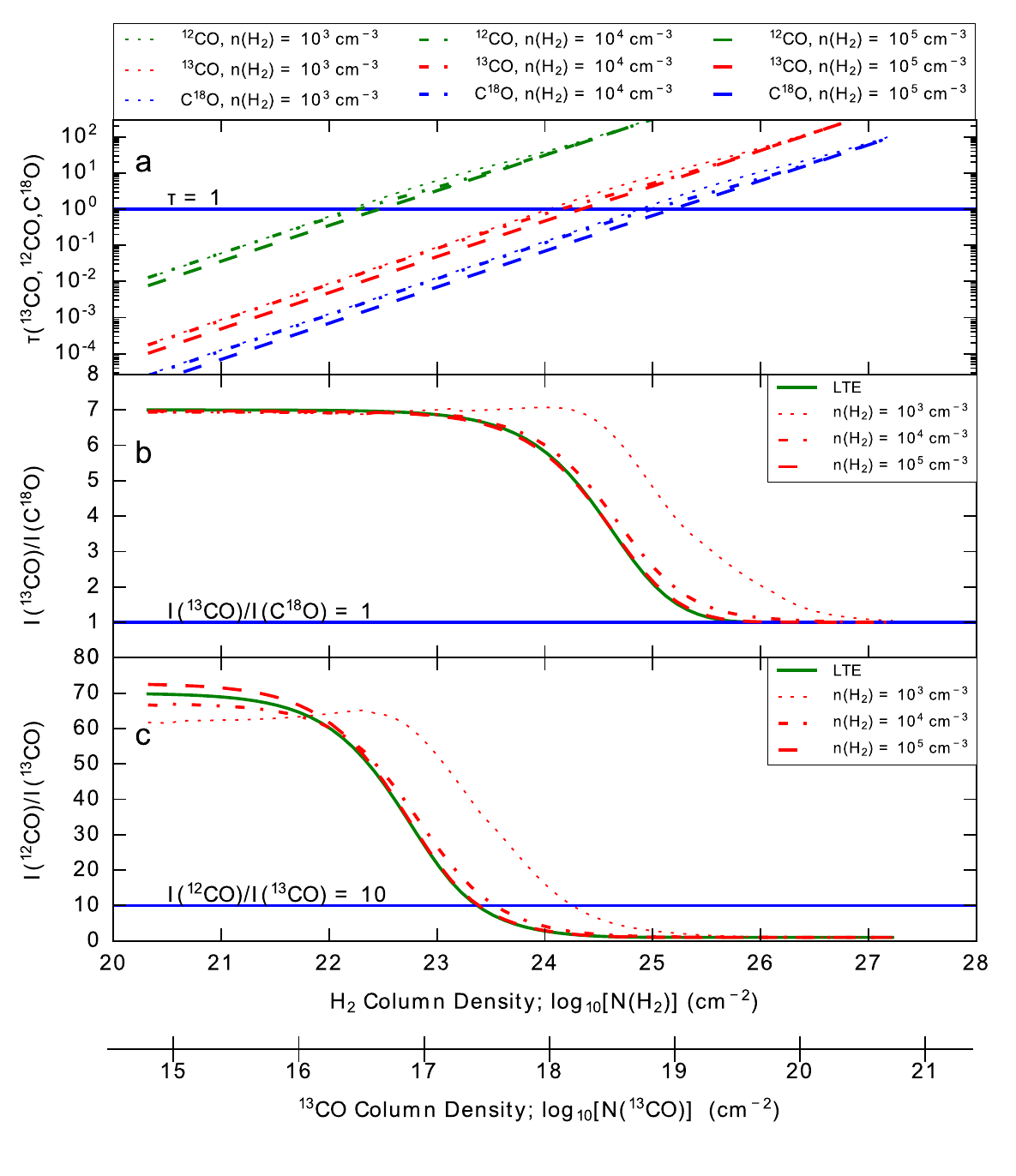}
\caption{
\textbf{a:} Optical depths of $^{12}$CO, $^{13}$CO and  C$^{18}$O, for the
$J$=3$\rightarrow$2 transition; \textbf{b:} $I$($^{13}$CO)/$I$(C$^{18}$O) line
ratio, and \textbf{c:} $I$($^{12}$CO)/$I$($^{13}$CO) line ratio as a function
of H$_2$\ column density, $n_{\rm H_2}$, and $^{13}$CO\ column density in various physical
conditions, for non-LTE models calculated with RADEX\cite{vdTak2007}. For all
models, we set the abundance ratios of $^{12}$CO, $^{13}$CO\ and C$^{18}$O\ to
be Galactic: $^{12}$CO/$^{13}$CO\ = 70 and $^{13}$CO/C$^{18}$O\  = 7, which are
representative values of the Milky Way disk.  Different line styles show the
gas conditions of $n_{\rm H_2}$\ = $10^3$, $10^4$ and $10^5$ cm$^{-3}$. The
$T_{\rm kin}$\ for all models are set to 30\,K, which is a typical dust
temperature for the SMG population, and the lowest $T_{\rm kin}$\ that H$_2$\
gas can reach for such intensive starburst conditions, due to cosmic ray
heating\cite{PPP2014}. In panels \textbf{b} and \textbf{c}, we also overlay the
line ratio (in thick green lines) with the LTE assumption for comparison.  All
three plots show that for Galactic abundances the line ratio of
$^{13}$CO/C$^{18}$O\ can approach unity only when the $^{13}$CO\ column density
is higher than $10^{19}$--$10^{20}$ cm$^{-2}$ (i.e.\ H$_2$ column density $N_{\rm
H_2}$\ $>$ $10^{25}$--$10^{26}$ cm$^{-2}$).} \label{fig:tau}
\end{figure*}

\begin{figure}
\centering
\includegraphics[scale=1.0]{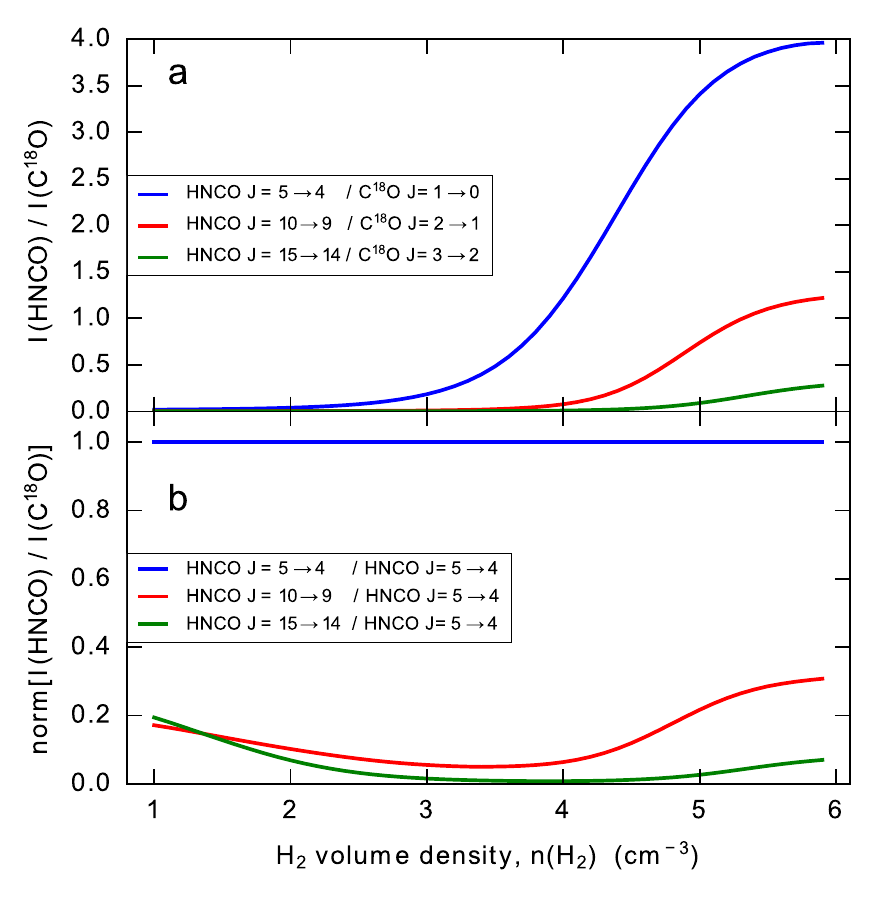}
\caption{
\textbf{a:} $I$(HNCO)/$I$(C$^{18}$O) line ratio as a function of H$_2$\ volume
density.  \textbf{b:} $I$(HNCO)/$I$(C$^{18}$O) line ratio as a function of H$_2$\
volume density, normalised with $I$(HNCO $J=5$$\rightarrow$\,4)/$I$(C$^{18}$O\ $J=1$$\rightarrow$\,0).
Both ratios are calculated using RADEX\cite{vdTak2007}, in which we assume the
same abundances as measured in Arp\,220 \cite{Martin2009}.  We assume $T_{\rm kin}$\ =
30 K as the representative kinetic temperature of the H$_2$\ gas.}
\label{fig:hnco}
\end{figure}
\clearpage

\begin{table*}[tbh]
\caption{ALMA Observational information} \label{tab:observations}
\vspace{5mm}
\begin{center}
\fontsize{7pt}{10}\selectfont
\begin{tabular}{cccccccc}\hline
Target                    & SPT\,0103$-$45         & SPT\,0103$-$45         & SPT\,0125$-$47         & SPT\,0125$-$47         & SDP.17B                & SDP.17B                & Cloverleaf \\
Isotopologue              & $^{13}$CO, C$^{18}$O\  & $^{12}$CO              & $^{13}$CO, C$^{18}$O\  & $^{12}$CO\             & $^{13}$CO, C$^{18}$O\  & $^{13}$CO, C$^{18}$O\  & $^{13}$CO, C$^{18}$O\  \\
Transition                & $J=5$$\rightarrow$\,4  & $J=5$$\rightarrow$\,4  & $J=3$$\rightarrow$\,2  & $J=4$$\rightarrow$\,3  & $J=3$$\rightarrow$\,2  & $J=4$$\rightarrow$\,3  & $J=3$$\rightarrow$\,2 \\
\hline
Observing Date            & 21-Jan-2016            & 21-Jan-2016            & 21-Jan-2016            & 21-Jan-2016            & 17-Jan-2016            & 16-Jan-2016            & 08-Apr-2016            \\
Bandpass Calibrator       & J2357$-$5311           & J2357$-$5311           & J2357$-$5311           & J2357$-$5311           & J0854+2006             & J0854+2006             & J1337$-$1257             \\
Flux Calibrator           & Neptune                & J2357$-$5311           & Neptune                & Neptune                & J0854+2006             & J0854+2006             & Callisto               \\
Gain Calibrator           & J0056$-$4451           & J0051$-$4226           & J0124$-$5113           & J0124$-$5113           & J0909+0121             & J0909+0121             & J1415+1320             \\
Integration Time (s)      & 1300                   & 120                    & 605                    & 120                    & 816                    & 726                    & 1753                   \\
Median PWV (mm)           & 6.1                    & 5.6                    & 6.2                    & 6.0                    & 2.0                    & 3.2                    & 3.0                    \\
Median T$_{\rm{sys}}$ (K) & 86                     & 85                     & 93                     & 88                     & 52                     & 60                     & 70                     \\
Angular Resolution        & 2.5$''$$\times$1.8$''$ & 2.7$''$$\times$1.6$''$ & 3.6$''$$\times$2.5$''$ & 2.5$''$$\times$1.8$''$ & 3.1$''$$\times$2.3$''$ & 2.7$''$$\times$1.8$''$ & 3.7$''$$\times$2.0$''$ \\
\hline
\end{tabular}\\
\end{center}
\noindent PWV, precipitable water vapour.
\end{table*}

\begin{table*}[tbh]
\caption{Observed targets, lines, frequencies, line widths and fluxes.}
\vspace{5mm}
\label{tab:fluxes}
\begin{center}
\fontsize{7pt}{10}\selectfont
\begin{tabular}{llllllllll}\hline
        Target    & Transition                     & $\nu_{\rm obs}$ & $I_{\rm line}$       & $I_{\rm line}^{\rm mom0}$ & $\sigma^{\rm theo}$$\star$ & $\Delta\, V_{\rm line}$ & $F_{\rm peak}$ \\ 
                  & $J\rightarrow J-1$             & GHz             & Jy\,km\,s$^{-1}$     & Jy\,km\,s$^{-1}$      & Jy\,km\,s$^{-1}$           & km\,s$^{-1}$            & mJy            \\ 
\hline
SDP.17b           & $^{12}$CO\ $J= 4\rightarrow3$  & 139.49          & $9.1 \pm 0.3 $       & --                    & --                         & 320                     & $\sim$40       \\ 
SDP.17b           & $^{13}$CO\ $J= 3\rightarrow2$  & 100.02          & $0.32\pm 0.05$       & $ 0.34\pm 0.08 $      & $0.08$                     &                         & $0.9\pm 0.3$   \\ 
SDP.17b           & C$^{18}$O\  $J= 3\rightarrow2$ & 99.64           & $0.26\pm 0.05$       & $ 0.32\pm 0.08 $      & $0.08$                     &                         & $0.8\pm 0.3$   \\ 
SDP.17b           & $^{13}$CO\ $J= 4\rightarrow3$  & 133.36          & $0.47\pm 0.07$       & $ 0.46\pm 0.08 $      & $0.07$                     &                         & $1.3\pm 0.4$   \\ 
SDP.17b           & C$^{18}$O\  $J= 4\rightarrow3$ & 132.85          & $0.34\pm 0.06$       & $ 0.50\pm 0.08 $      & $0.07$                     &                         & $1.0\pm 0.4$   \\ 
\hline
Cloverleaf        & $^{12}$CO\ $J= 3\rightarrow2$  & 97.17           & $13.2\pm 0.2$        & --                    & --                         & 400                     & $30 \pm 1.7$   \\ 
Cloverleaf        & $^{13}$CO\ $J= 3\rightarrow2$  & 92.90           & $0.65\pm 0.09$       & $ 0.61\pm 0.06 $      & $0.07$                     &                         & $1.4\pm 0.4$   \\ 
Cloverleaf        & C$^{18}$O\  $J= 3\rightarrow2$ & 92.55           & $0.40\pm 0.10$       & $ 0.43\pm 0.06 $      & $0.07$                     &                         & $0.8\pm 0.4$   \\ 
\hline
SPT\,0103$-$45    & $^{12}$CO\ $J= 4\rightarrow3$  & 112.68          & $8.2 \pm 0.6$        & --                    & --                         &                         & $32 \pm 0.6$   \\ 
SPT\,0103$-$45    & $^{12}$CO\ $J= 5\rightarrow4$  & 140.91          & $8.8 \pm 0.5\dagger$ & $8.8\pm 0.6\dagger$   & --                         & 300$^\dagger$           & $27.8 \pm0.2$  \\ 
SPT\,0103$-$45    & $^{13}$CO\ $J= 5\rightarrow4$  & 134.65          & $0.37\pm0.07$        & $0.38\pm0.05 $        & $0.07$                     &                         & $1.2 \pm 0.4$  \\ 
SPT\,0103$-$45    & C$^{18}$O\  $J= 5\rightarrow4$ & 134.13          & $0.35\pm0.09$        & $0.39\pm0.07 $        & $0.07$                     &                         & $1.2 \pm 0.4$  \\ 
\hline
SPT\,0125$-$47    & $^{12}$CO\ $J= 3\rightarrow2$  & 98.38           & $18.1\pm 0.5$        & $18.0\pm 0.5$         & --                         &                         & $43 \pm 4$     \\ 
SPT\,0125$-$47    & $^{12}$CO\ $J= 4\rightarrow3$  & 131.21          & $26.9\pm 0.7$        & $26.8\pm 0.7$         & --                         & 400                     & $69 \pm 3$     \\ 
SPT\,0125$-$47    & $^{13}$CO\ $J= 3\rightarrow2$  & 94.06           & $0.78\pm 0.09$       & $0.86\pm 0.07$        & $0.1$                      &                         & $2.0 \pm 0.4$  \\ 
SPT\,0125$-$47    & C$^{18}$O\  $J= 3\rightarrow2$ & 93.70           & $0.63\pm 0.07$       & $0.71\pm 0.1 $        & $0.1$                      &                         & $1.6 \pm 0.4$  \\ 
\hline\\
\end{tabular}\\
\end{center}
\noindent
Literature data are from refs \cite{Omont2013,Weiss2003,Weiss2013}. $T_{\rm sys}$ is the system temperature.
$\star$ Theoretical noise level calculated using the ALMA
sensitivity calculator\cite{ALMAcalculator}.\\
$\dagger$ There are two velocity components in the $^{12}$CO\ spectrum. We adopt only
the narrow component, seen for $^{12}$CO\ $J=5\rightarrow4$, to avoid the broad and
weaker component (see Extended Data Fig.~\ref{fig:spectra}).
\end{table*}

\end{extendeddata}

\clearpage


\begin{addendum}
\item 
The authors are grateful to the referees for their constructive suggestions and
comments. Z.-Y.Z. is grateful to Xiaoting Fu, Hau-Yu Baobab Liu, Yancy Shirley,
and Peter Barnes for helpful discussions.  Z.-Y.Z., R.J.I.\ and P.P.P.\
acknowledge support from the European Research Council in the form of the
Advanced Investigator Programme, 321302, COSMICISM. F.M. acknowledges financial
funds from Trieste University, FRA2016. This research was supported by the
Munich Institute for Astro- and Particle Physics (MIAPP) of the DFG cluster of
excellence \textit{``Origin and Structure of the Universe''}.  This work also
benefited from the International Space Science Institute (ISSI) in Bern, thanks
to the funding of the team \emph{``The Formation and Evolution of the Galactic
Halo''} (PI D.~Romano) This paper makes use of the ALMA data. ALMA is a
partnership of ESO (representing its member states), NSF (USA) and NINS
(Japan), together with NRC (Canada), MOST and ASIAA (Taiwan), and KASI
(Republic of Korea), in cooperation with the Republic of Chile. The Joint ALMA
Observatory is operated by ESO, AUI/NRAO and NAOJ.

\item[Author Contributions ] 
Z.-Y.Z.\ is the Principal Investigator of the ALMA observing project. Z.-Y.Z.
reduced the data and wrote of the initial manuscript.  R.J.I.\ and P.P.P.\
provided ideas to initialise the project and helped write the manuscript.
Z.-Y.Z. and P.P.P. worked on molecular line modeling of isotopologue ratios and
chemical/thermal effects on the abundances. D.R.\ and F.M.\ ran the chemical
evolution models and provided theoretical interpretation of the data. All
authors discussed and commented on the manuscript.

\item[Competing interests]
The authors declare no competing financial interests.

\item[Code Availability]

We opt not to make the code used for the chemical evolution modeling publicly
available because it is an important asset of the researchers' toolkits.  The
code for analysing the line ratios and optical depths of $^{12}$CO, $^{13}$CO,
C$^{18}$O\ are based on the publicly available non-LTE radiative transfer code,
RADEX\cite{vdTak2007}.

\item[Data Availability ]

The dataset that supports the findings of this study is available in the ALMA
archive (http://almascience.eso.org/aq/) under the observing project
\#2015.1.01309.S,  \#2013.1.00164.S,  \#2011.0.00958.S, and \#2011.0.00747.S. 
Additional requests can be directed to the corresponding author upon reasonable
request.

\end{addendum}

\end{document}